\documentclass{ws-procs975x65}

\begin{document}

\title{THE GENERALIZED SECOND LAW AND THE EMERGENT UNIVERSE}
\author{D. PAV\'{O}N$^*$}
\address{Departamento de F\'{\i}sica, Universidad Aut\'{o}noma de Barcelona,\\
Bellaterra, 08193, Spain\\
$^*$E-mail: diego.pavon@uab.es}
\author{S. DEL CAMPO and R. HERRERA}
\address{Instituto de F\'{\i}sica, Pontificia Universidad
Cat\'{o}lica de Valpara\'{\i}so, Chile}
\begin{abstract}
We explore whether the generalized second law of thermodynamics is
fulfilled in the transition from a generic initial Einstein static
phase to the inflationary phase, with constant Hubble rate, and
from the end of the latter to the conventional thermal radiation
dominated era of expansion. As it turns out, the said law is
satisfied provided the radiation component does not contribute
largely to the total energy of the static phase.
\end{abstract}


\bodymatter

\section{Introduction}
\label{sec:intro}
Different cosmological scenarios have been devised to evade the
initial singularity of the big bang standard model. These include
bouncing universes and the emergent universe. Here we shall focus
on a representative of the latter put forward by Ellis and
Maartens \cite{cqg-george}. In this scenario  the initial
singularity is replaced by an Einstein static phase in which the
scale factor of the Friedmann-Robertson-Walker metric does not
vanish and, accordingly, the energy density, pressure and so on
stay finite. Thus, the Universe starts expanding from the said
phase, smoothly joins a stage of exponential inflation followed by
standard reheating to subsequently approach the classical thermal
radiation dominated era of the conventional big bang cosmology.
Figure \ref{aba:fig1} depicts this evolution. Fairly generally,
the static phase is dominated pressureless matter and radiation
(subscripts $m$ and $\gamma$, respectively), curvature (which has
to be positive, $k = +1$) and a scalar field. Thus the total
energy density in this phase (subscript $I$) obeys, $\rho_{m,I}\,
+ \, \rho_{\gamma,I} \, + (1/2)\dot{\phi}^{2}_{I} \, + V_{I} =
\frac{3k}{8 \pi G a_{I}^{2}}$. In this scenario the potential must
be asymptotically flat in the infinite past,
\begin{equation}
V(\phi) \rightarrow V_{I} \quad {\rm as} \quad \phi \rightarrow
-\infty\, , \quad t \rightarrow - \infty \, , \label{eq:Vflat}
\end{equation}
and fall toward a minimum at some finite value. Accordingly, the
field rolls down from the static state at $- \infty$ and the
potential slowly decreases from its initial value, $V_{I}$, in the
infinite past. To have acceleration the inequality $V(\phi) -
\dot{\phi}^{2} > 0$  ought to be fulfilled. Since $V(\phi)$
decreases and $\dot{\phi}^{2}$ augments, at some time, say $t=
t_{e}$, inflation terminates, then $\phi$ oscillates about the
minimum and reheating takes place, the latter followed by the
radiation dominated era.

As demonstrated by  Bekenstein, the entropy of a black hole plus
the entropy of its surroundings cannot diminish \cite{jakob}. This
law, aptly named the ``generalized second law" (GSL) of
thermodynamics as it considers conventional matter/fields and an
event horizon, was extended by several authors to cosmological
settings in which the black hole horizon is replaced by a causal
cosmic horizon \cite{gsl}. This version of the said law
establishes that the entropy of the horizon plus the entropy of
the matter and fields within the horizon can never decrease. In
this note we study which constraints (if any) the GSL imposes on
the two intermediate phases, i.e., from the static phase to
exponential inflation and from the reheating to thermal radiation
domination (in the static, inflationary, and thermal radiation
dominated phases the GSL is trivially fulfilled; see \cite{plb}
for details). As causal horizon we shall consider the apparent
horizon of area ${\cal A} = 4 \pi \tilde{r}_{A}^{2}$ where
$\tilde{r}_{A} = 1/\sqrt{H^{2}+ka^{-2}}$ denotes its radius
\cite{bak}. Notice that neither the particle horizon nor the event
horizon exist in the static phase; only the particle horizon is
meaningful in all the phases considered here. Neglecting quantum
effects, the horizon entropy can be written as $S_{{\cal A}} =
k_{B} \, \frac{{\cal A}}{4\, \ell^{2}_{pl}}$. Assuming the scalar
field is in a pure quantum state, the GSL reads $S'_{{\cal A}} \,
+ \, S'_{m}\, + \, S'_{\gamma} \geq 0$, where the prime means
derivative with respect the scale factor, $a$.
\section{The GSL at the transitions phases}
\label{sec:trans}
\begin{figure}[t]
\begin{center}
\psfig{file=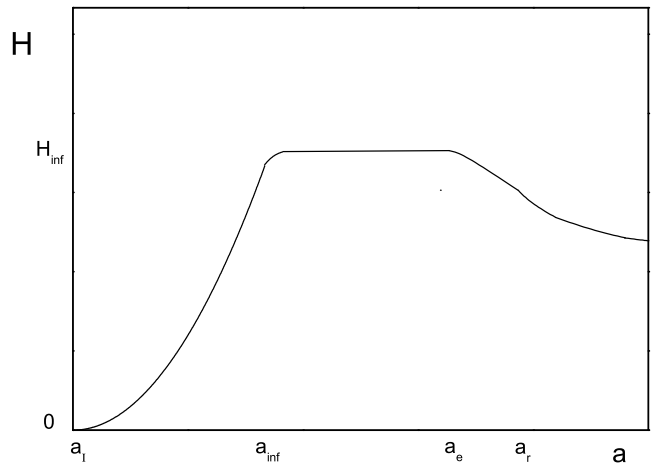,width=4in}
\end{center}
\caption{Schematic evolution of the Hubble function from the
Einstein static era to the thermal radiation era. Here $a_{inf}$
and $a_{e}$ stand for the scale factor at the beginning and end of
exponential inflation, respectively; $a_{r}$ denotes the scale
factor at some generic point at the radiation dominated expansion
era.} \label{aba:fig1}
\end{figure}

In the transition from the static to the inflationary phase
($a_{I} < a < a_{inf}$, see Fig.~\ref{aba:fig1}) one has $
S'_{{\cal A}}= k_{B}/(2 \ell^{2}_{pl})H {\cal A}^{2}(\rho + p)/a >
0$, as well as
\begin{equation}
S'_{m} = - 3 k_{B} \, \frac{N}{a_{I}^{3}}\,\tilde{r}_{A}^{5} \,
\left(H \, H' \, - \, \frac{k}{a^{3}}\right)\, , \quad T_{\gamma}
\, S'_{\gamma} = 2 \pi \, (1\, + \, w_{\gamma}) (1 \, + \, 3w)
\tilde{r}_{A}^{3} \frac{\rho_{\gamma}}{a}\, , \label{eq:Sderiv}
\end{equation}
where $w_{\gamma} = p_{\gamma}/\rho_{\gamma}$ and $w = p/\rho$ is
the equation of state parameter of the overall fluid -including
the scalar field.

With the help of the Einstein field equation
\begin{equation}
H \, H' \, - \, \frac{k}{a^{3}} = -4 \pi \, G\, \frac{\rho \, + \,
p}{a} \, , \label{eq:Hprimegr}
\end{equation}
it follows that the GSL is fulfilled provided the upper bound
\begin{equation}
\frac{\rho_{\gamma}}{\rho} \leq \frac{3}{4}\, \frac{k_{B} G \pi
(1\, + \, w) T_{\gamma I} \, a_{I} \left[\frac{4}{\ell^{2}_{pl}}
\, + \, 6 \frac{N}{a_{I}^{2}}\right]}{|1\, + \, 3w|}\, ,
\label{constraint2}
\end{equation}
on the amount of radiation energy in the stationary phase is met
-see \cite{plb} for details.

The transition between the period of exponential inflation and the
thermal radiation dominated phase begins at $a = a_{e}$ and ends
when the products of the inflaton decay at reheating get fully
thermalized. In this phase, $H' < 0$, the pressureless matter
particles have essentially disappeared, and again $S'_{\cal{A}} >
0$. In its turn, the entropy of the mixture of radiation and
relativistic particles originated in the decay of the inflaton is
given by Eq. (\ref{eq:Sderiv}.2), with $w$ and $w_{\gamma}$
replaced by $\tilde{w}$ which is positive-definite. (The tilde is
to reminds us that the mixture is not thermalized though
$\tilde{w} \rightarrow w_{\gamma} = 1/3$ as the thermalization
process goes on). Thus, in this transition the GSL is guaranteed.
\section{Conclusions}
For a cosmological model to be worthy of consideration, aside from
passing the observational tests, it must comply with
thermodynamics; more specifically, it must respect the GSL. We
have shown that the toy model of Ref.~\cite{cqg-george} is
thermodynamically safe provided that the radiation energy does not
contribute largely to the static phase. Finally, it should be
explored if this nice feature is also present in other emergent
scenarios.
\section*{Acknowledgments}
This work was partially supported by the Chilean grant FONDECYT
N$_{0}$ 1110230.


\end{document}